\documentclass[letterpaper,12pt]{article}
\usepackage{tabularx}
\usepackage{amsmath}  
\usepackage{graphicx,xcolor}
\usepackage[margin=1in,letterpaper]{geometry}
\usepackage{cite} 
\usepackage{physics}
\usepackage{tensor}
\usepackage{fancyhdr}
\usepackage{authblk,comment}

\usepackage{jheppub}
\title{Holographic D-brane constructions with dynamical gauge fields}

\author[a,b,c]{Yongjun Ahn,}
\author[a,b,c]{Matteo Baggioli,}
\author[d]{Hyun-Sik Jeong,}
\author[e]{Masataka Matsumoto}

\affiliation[a]{School of Physics and Astronomy, Shanghai Jiao Tong University, Shanghai 200240, China}
\affiliation[b]{Wilczek Quantum Center, School of Physics and Astronomy, Shanghai Jiao Tong University, Shanghai 200240, China}
\affiliation[c]{Shanghai Research Center for Quantum Sciences, Shanghai 201315, China}
\affiliation[d]{Instituto de F\'isica Te\'orica UAM/CSIC, Calle Nicol\'as Cabrera 13-15, 28049 Madrid, Spain}
\affiliation[e]{Department of Physics, Chuo University, Tokyo 112-8551, Japan}

\emailAdd{yongjunahn@sjtu.edu.cn}
\emailAdd{b.matteo@sjtu.edu.cn}
\emailAdd{hyunsik.jeong@csic.es}
\emailAdd{mmatsumoto173@g.chuo-u.ac.jp}

\preprint{\texttt{IFT-UAM/CSIC-25-47}}

\abstract{Holographic D-brane constructions, governed by the Dirac-Born-Infeld (DBI) action, play a central role in the AdS/CFT correspondence, particularly in applications to quantum chromodynamics and condensed matter systems. In this work, we demonstrate how to equip these bottom-up holographic models with dynamical boundary gauge fields, thereby introducing electromagnetic interactions into their dual field theory descriptions. As a direct application of this formalism, we compute the dispersion relations of the lowest quasinormal modes around both equilibrium and nonequilibrium steady states, and show that their behavior matches the predictions from hydrodynamics with dynamical $U(1)$ symmetry.
}

\begin{document}

\maketitle

\section{Introduction} \label{sec:intro}
From a conceptual standpoint, nonequilibrium steady states (NESS) are the closest analogs to thermal equilibrium within the broader class of nonequilibrium phenomena. NESS arise when external driving forces are precisely balanced by dissipation through steady currents, resulting in macroscopic quantities that remain constant in time. Classic examples include the Drude model of electrical conductivity and steady-state heat flow in a thermal conductor.

Despite this resemblance to equilibrium, NESS violate foundational principles in statistical physics, such as the detailed balance condition. Consequently, conventional thermodynamic and hydrodynamic frameworks cannot be applied directly. To address this, several theoretical approaches have been developed, such as the matrix product ansatz~\cite{Blythe_2007}, the additivity principle~\cite{Bodineau_2004}, and macroscopic fluctuation theory~\cite{Bertini_2015}. More recently, efforts have been made to formulate thermodynamic descriptions of NESS~\cite{nakagawa2019global,Liu:2018crr,Kundu:2019ull}, and to extend hydrodynamic theories to capture their low-energy dynamics~\cite{Amoretti:2024jig,Amoretti:2022ovc,Brattan:2024dfv}.

To test and refine these theoretical developments, one needs microscopic models where non-equilibrium physics and relaxation dynamics can be computed reliably. In this context, the holographic correspondence offers a powerful framework. By mapping strongly coupled quantum systems to higher-dimensional classical gravity theories, holography has provided deep insights into many-body physics across fields ranging from quantum chromodynamics (QCD) to condensed matter theory (CMT) and quantum information~\cite{Hartnoll2016apf,Zaanen:2015oix,Baggioli:2019rrs,Natsuume:2014sfa}. In particular, thermodynamic and hydrodynamic behavior are encoded in black hole geometries and their quasi-normal modes (QNMs)~\cite{Horowitz:1999jd,Birmingham:2001pj,Son:2002sd,Kovtun:2005ev}.

A particularly fruitful holographic setup for studying NESS is the D3/D7 probe brane system~\cite{Karch:2002sh}. In this top-down construction, D3-branes source the background geometry dual to ${\cal{N}}=4$ supersymmetric Yang-Mills theory, while probe D7-branes introduce ${\cal{N}}=2$ flavor degrees of freedom. By applying a constant electric field to the D7-brane, one can drive a steady current through the system, realizing an electrically driven NESS~\cite{Karch:2007pd,Nakamura:2010zd}. Dissipation occurs via energy transfer into the gravitational thermal bath. This model provides a concrete and tractable example of an NESS in a strongly coupled setting. Recent studies have demonstrated that the QNM spectrum of such systems exhibits features consistent with relaxed hydrodynamics, especially in the slow relaxation regime~\cite{Brattan:2024dfv}.

Another key ingredient for realistic modeling is the inclusion of long-range electromagnetic field. In particular, relativistic magnetohydrodynamics (MHD) describes the interplay between charged fluids and dynamical electromagnetic fields, which gives rise to phenomena such as Alfv\'{e}n and magnetosonic waves~\cite{Hernandez:2017mch,Lier:2025wfw}. In holography, this can be achieved by imposing ``mixed boundary conditions" for the bulk gauge fields. Following Witten’s seminal proposal~\cite{Witten:1998qj,Witten:2003ya} and its extensions~\cite{Klebanov:1999tb,Leigh:2003ez,Yee:2004ju,Breitenlohner:1982jf,Marolf:2006nd}, this procedure corresponds to adding boundary terms, such as double-trace deformations, that induce a Maxwell kinetic term for the boundary gauge field. Physically, this allows one to incorporate electric screening and propagating electromagnetic modes in dual field theory.

Mixed boundary conditions have been widely applied in holography to study a range of phenomena, including plasmons~\cite{Gran:2017jht,Gran:2018iie,Baggioli:2019aqf,Baggioli:2019sio,Baggioli:2021ujk,Mauri:2018pzq,Romero-Bermudez:2018etn}, Friedel oscillations~\cite{Faulkner:2012gt}, anyons~\cite{Jokela:2013hta,Brattan:2013wya,Brattan:2014moa}, relativistic MHD~\cite{Ahn:2022azl}, and superconductivity with Meissner effects and Anderson-Higgs mechanisms~\cite{Jeong:2023las,Baggioli:2023oxa,Natsuume:2022kic,Natsuume:2024sic,Natsuume:2024ril}. Related methods have also been used to promote the boundary metric to a dynamical variable, leading to semiclassical Einstein equations in the dual theory~\cite{Compere:2008us,Ecker:2021cvz,Ishibashi:2023luz}.

In this work, we extend the framework of relaxed hydrodynamics around current-driven NESS by incorporating a dynamical electromagnetic field. To test the validity of this extension, we analyze the QNM spectrum of a deformed D3/D7 model with mixed boundary conditions for the bulk gauge field, and subjected to a finite electric field, and compare the results with hydrodynamic predictions in the slow relaxation regime. More broadly, we introduce dynamical boundary gauge fields in D-brane constructions, with potential applications ranging from AdS-CMT to AdS-QCD frameworks.

%
\section{Relaxed hydrodynamic theory with dynamical gauge fields}\label{sec2}

In this section, we develop a hydrodynamic framework to describe electrically driven, nonequilibrium steady states of charged fluids in $(3+1)$ dimensions, explicitly incorporating dynamical electromagnetic effects. Building upon previous work~\cite{Brattan:2024dfv} (see also \cite{Amoretti:2024jig,Amoretti:2022ovc}), we extend the model by including a finite electromagnetic coupling. To illustrate our approach, we focus on systems characterized by a uniform charge density $\rho$ and a stationary charge current $\vec{J} = \sigma_{\text{DC}}(\rho, \vec{E}^2) \vec{E}$, where $\sigma_{\text{DC}}$ denotes the nonlinear direct current (DC) conductivity.

For simplicity and clarity, we adopt the ``probe limit", meaning we ignore fluctuations in the stress tensor. This approximation is valid when the environment or bath has significantly more degrees of freedom compared to the system under study.

\paragraph{Equations of motion and constitutive relation.}
Our primary interest is then in fluctuations described by the charge conservation equation:
\begin{equation}
\label{eq:currcon}
    \partial_{t} \delta\rho  + \partial_{i} \delta J^{i} = 0\,,
\end{equation}
where $\delta \rho$ represents a linearized fluctuation of the charge density, and $\delta J^i$ corresponds to fluctuations in the spatial current.
In addition, we also include Maxwell's equations to capture dynamical electromagnetism associated with electric ($\delta \vec{E}$) and magnetic ($\delta \vec{B}$) field fluctuations:
\begin{equation}
\begin{split}
\label{eq:Maxwell}
0&=\epsilon_e \nabla\cdot \delta \vec{E}-\delta\rho\,, \qquad\qquad\,\,\,\,\, 
0=\nabla\cdot \delta \vec{B}\,,\\
0&=\nabla\times\delta \vec{E}+\partial_t\delta\vec{B}\,,\qquad\qquad
0=\nabla\times\delta \vec{B}-\mu_m\epsilon_e\partial_t \delta \vec{E}-\mu_m \delta\vec{J}\,,
\end{split}
\end{equation}
with $\epsilon_e$ and $\mu_m$ being the electric permittivity and magnetic permeability, respectively.

To complete our hydrodynamic description, i.e., to solve equations of motion \eqref{eq:currcon}-\eqref{eq:Maxwell}, we specify a constitutive relation for the current fluctuations. To first-order approximation and at small electric field amplitudes (compared to other scales in the system, like the charge density), this relation becomes~\cite{Chen:2017dsy,Brattan:2024dfv}:
\begin{equation}
\label{eq:currentflucdyn}
\partial_{t} \delta J^{i} + \partial_{j} \mathcal{T}^{ij} - \alpha E^i\delta \rho  + \frac{1}{\tau} \delta J^{i} = \chi\delta E^{i} \,,
\end{equation}
where
\begin{align}
   \mathcal{T}^{ij}=\,&v^{2} \delta^{ij} \delta \rho + \left( \eta_{1} + \eta_{2} \right) \frac{E^{i} E^{j}  E_{k}}{\vec{E}^{2}} \delta J^{k}  + \left( \eta_{2} +\eta_{3} \right) E^{(i} \Pi\indices{_{k}^{j)}} \delta J^{k}\nonumber\\
   &+ \left( \eta_{2} - \eta_{3} \right) E^{[i} \Pi\indices{_{k}^{j]}} \delta J^{k}  + \eta_{4} \Pi\indices{^{ij}} \delta J^{k} E_{k} \vphantom{\frac{E^{i} E^{j}  E_{k}}{\vec{E}^{2}}} \,.\label{lala}
\end{align}
Here, $\Pi^{ij} = \delta^{ij} - \frac{E^i E^j}{\vec{E}^2}$ is a projector with respect to the electric field and $\{v, \alpha, \tau, \chi, \eta_i \}$ are associated with various transport coefficients~\cite{Amoretti:2023vhe,Brattan:2024dfv}, for instance, $v$, the speed of sound, is related to the diffusion constant $D=\tau v^{2}$ and $\alpha$ with the drift velocity $v_\mathrm{drift}^{i}$ in \eqref{DRIFVEL}, and $1/\tau$ is the decay rate of the weakly non-conserved $J$.

\paragraph{Connection to the heat conduction equation.}
Notably, our hydrodynamic equation \eqref{eq:currentflucdyn} resembles the Drude model of electron transport, containing terms for driving forces ($\sim E$), dissipation ($\sim 1/\tau$), and electromagnetic coupling through the current-current susceptibility $\chi$. After some manipulation of Eqs. \eqref{eq:currcon}-\eqref{eq:currentflucdyn}, the charge fluctuations satisfy an equation similar to the Cattaneo-Christov heat conduction model, highlighting a universal structure
\begin{equation}
	 \tau \frac{\partial^2 \delta \rho}{\partial t^2} + \frac{\partial \delta \rho}{\partial t} + \vec{v}_\mathrm{drift} \cdot \nabla \delta \rho - D \nabla^2 \delta \rho + K \delta \rho = \mathcal{R}, \label{eq:chargefluc}
\end{equation}
where
\begin{align}\label{ttt}
\begin{split}
\mathcal{R} &= \tau \eta_{4} \nabla_{\perp}^2 ( \vec{E} \cdot \delta \vec{J}) + \tau \left( \eta_{1} + \eta_{2} \right) \frac{( \vec{E} \cdot \nabla)^2}{\vec{E}^{2}} \vec{E} \cdot \delta \vec{J}+ \tau \left( \eta_{2} + \eta_{3} \right) (\vec{E} \cdot \nabla) (\nabla_{\perp} \cdot \delta \vec{J} ) \,,\\
K &= \tau \frac{\chi}{\epsilon_{e}} \,.
\end{split}
\end{align}
Here, we clarify the origin and physical interpretation of each term in Eq. \eqref{eq:chargefluc}. Specifically, the drift velocity term arises directly from the drift current $\alpha E^i \delta \rho$ in Eq. \eqref{eq:currentflucdyn}. The diffusion term originates from the isotropic pressure-like term $v^{2}\delta^{ij}\delta \rho$ in the stress tensor $\mathcal{T}^{ij}$. The parameter $K$ emerges due to electromagnetic interactions, reflecting how electromagnetic coupling influences charge dynamics, and $\left(\nabla_{\perp}\right)^i = \Pi^{ij} \nabla_j$ denotes the spatial derivative perpendicular to the electric field.

Interestingly, Eq. \eqref{eq:chargefluc} structurally resembles the heat conduction equation described by the Cattaneo-Christov model~\cite{Christov:2005aa,CHRISTOV2009481,Bissell_2015,STRAUGHAN201095}. In this analogy, the drift velocity plays the role of the fluid velocity in heat transport, the diffusion constant $D$ corresponds to thermal diffusivity, and the external source term $\mathcal{R}$ encapsulates additional driving forces or effects. However, a notable difference is that here the electric field couples directly to the charge density, unlike temperature, which couples through a time derivative in heat conduction.

This parallel indicates a broader universality in the hydrodynamic descriptions of nonequilibrium steady states, highlighting that despite distinct physical contexts, similar mathematical structures underpin various transport phenomena~\cite{Brattan:2024dfv}.

\paragraph{Dispersion relations and hydrodynamic modes.}
Next, analyzing our equations \eqref{eq:currcon}-\eqref{eq:currentflucdyn} in Fourier space ($\partial_\mu \to i k_\mu=\{-\omega, \vec{k}\}$), we derive dispersion relations for both longitudinal (parallel) and transverse (perpendicular) modes relative to the electric field. 
To begin with, one can find the relevant equations of motion as 
\begin{equation}
\begin{split}
0&=i \epsilon_e \,\vec{k}\cdot \delta\vec{E}-\delta \rho\,,\\
0&=i\vec{k}\cdot \delta \vec{B}\,,\\
0&=\frac{i}{\mu_m}\vec{k}\times \delta \vec{B}+i\epsilon_e\,\omega\,\delta\vec{E}-\delta\vec{J}\,,\\
0&=-i\omega\,\delta\vec{J}+i v^2 \vec{k}\,\delta \rho+\frac{1}{\tau}\delta{\vec{J}}-\chi \delta E - \alpha \vec{E} \delta\,{\rho}\\
&+i(\eta_1+\eta_2)\vec{k}\left(\vec{E}\cdot\delta \vec{J}_\parallel\right)+i\frac{\eta_2+\eta_3}{2}\left(\vec{E}\cdot\vec{k}\right)\delta\vec{J}_\perp-i\frac{\eta_2-\eta_3}{2}\left(\vec{E}\cdot\vec{k}\right)\delta\vec{J}_\perp\,,
\end{split}
\end{equation}
where the symbol $\parallel(\text{or} \perp)$ denotes the component of $\delta \vec{J}$ in parallel (or perpendicular) to the electric field. 

After appropriate manipulations, the system of equations yields two decoupled dispersion relations corresponding to longitudinal and transverse modes of the charge fluctuations.
For the longitudinal mode, the dispersion relation takes the form
\begin{equation}
\label{eq:hydrolongi}
0=1-\frac{\epsilon_e}{\chi}\left( \omega\left(\omega +\frac{i}{\tau}\right)-v^2k^2-i\alpha E k - E\left(\eta_1+\eta_2\right)\omega k \right)\,,
\end{equation}
and for the transverse mode
\begin{equation}
\label{eq:hydrotrans}
0=\omega\left(\frac{\chi}{\epsilon_e}-\frac{i}{\tau}\omega-\omega^2\right)+k\left(\eta_3 E\left(\omega^2-\frac{k^2}{\epsilon_e\mu_m}\right)+ \left(\omega+\frac{i}{\tau}\right)\frac{k}{\epsilon_e\mu_m}\right)\,.
\end{equation}
Here, to simplify the analysis, we introduce the identification:
\begin{equation}\label{eq:c2}
\epsilon_e=\frac{1}{\lambda}\,, \qquad \mu_m= \lambda\,,
\end{equation}
where $\lambda$ represents the electromagnetic coupling, ensuring the speed of light is normalized via $\epsilon_e \mu_m = 1$. This substitution allows us to examine how the electromagnetic coupling $\lambda$ impacts the system's collective excitations, particularly in the small wave-vector limit, where analytic expansions become tractable and physically insightful.

\paragraph{\quad (I) Longitudinal modes.}
Expanding the longitudinal dispersion relation from  Eq.\eqref{eq:hydrolongi} at small $k$, we find two modes
\begin{align}\label{LONGIDIS}
\omega_{\pm}^{\text{L}} = -\frac{i}{2\tau} \left( 1 \pm \sqrt{1-4 \tau^2 \chi  \lambda} \right) + \frac{1}{2} \left( \eta_1 + \eta_2 \pm \frac{\eta_1 + \eta_2 -2 \alpha \tau}{\sqrt{1 - 4\tau^2 \chi \lambda}} \right) {E}\,{k} \,.
\end{align}
In the absence of electric fields, the mode $\omega_{-}^{\text{L}}$ corresponds to the familiar damped charge diffusion mode with a relaxation rate set by $\tau^{-1}$~\cite{Ahn:2022azl}. When a finite electric field $E$ is present, the dispersion acquires a linear $k$-dependent term, capturing the drift of charge due to the field. Importantly, a non-zero electromagnetic coupling $\lambda$ influences both the damping rate and the drift velocity. If $\lambda > 1/(4\tau^2 \chi)$, the system even develops a real frequency gap, signaling a qualitative change in mode behavior.

\paragraph{\quad (II) Transverse modes.}
For the transverse sector, we expand the dispersion relation from Eq.\eqref{eq:hydrotrans} in the limit of small $k$ and $\lambda$. The result yields three distinct modes:
\begin{align}
\begin{split}\label{TRANIDIS}
\omega_{\text{gapless}}^{\text{T}} &= -i \frac{k^2}{\tau \chi \lambda} \,, \\
\omega_{\text{gap, (1)}}^{\text{T}} &= -i \tau \chi \lambda - \eta_3 \tau^2 \chi \lambda \, {E}\,{k} + i \left(\frac{1}{\tau \chi \lambda} + \left(1 + \eta_3^2 E^2\right) \tau^3 \chi \lambda  \right)k^2 \,, \\
\omega_{\text{gap, (2)}}^{\text{T}} &= -\frac{i}{\tau} \left(1-\tau^2 \chi \lambda \right) + \eta_3 \left( 1 + \tau^2 \chi \lambda \right) {E}\,{k}\,.
\end{split}
\end{align}
It is crucial to treat the $\lambda \to 0$ limit with care. At exactly $\lambda = 0$ in \eqref{eq:hydrotrans}, the first two modes revert to propagating electromagnetic waves with linear dispersion $\omega = \pm k$. However, for finite $\lambda$, these modes transform as: 
\begin{itemize}
\item{$\omega_{\text{gapless}}^{\text{T}}$ becomes a diffusive magnetic mode with diffusion constant $D_B = 1/(\tau \chi \lambda)$.}
\item{$\omega_{\text{gap,(1)}}^{\text{T}}$ describes the relaxation of electric field fluctuations with a characteristic rate $\tau_e^{-1} = \tau \chi \lambda$.}
\end{itemize}
These results reflect the underlying symmetry structure: while the electric $U(1)$ symmetry is explicitly broken by the field, magnetic flux remains conserved, giving rise to magnetic diffusion~\cite{Grozdanov:2016tdf, Ahn:2022azl}.

Moreover, these electromagnetic modes are fundamentally tied to the presence of dynamical gauge fields and are absent in frameworks that neglect such dynamics, such as in~\cite{Brattan:2024dfv}.

\paragraph{Electromagnetic coupling and drift velocity corrections.}
A particularly insightful result is the distinction between longitudinal and transverse drift velocities in the presence of finite electromagnetic coupling. While they coincide at $\lambda = 0$, a small but finite $\lambda$ induces corrections proportional to $\tau^3 \chi \lambda$, leading to:
\begin{align}
\begin{split}\label{DRIFVEL}
v^{\text{L}}_{\text{drift}} = \alpha \tau + \frac{2}{3} \alpha \tau^3 \chi \lambda \,, \qquad
v^{\text{T}}_{\text{drift}} = \alpha \tau +  \alpha \tau^3 \chi \lambda \,, 
\end{split}
\end{align}
which are read off from the linear-in-$k$ terms of $\omega_{-}^{\text{L}}$ and $\omega_{\text{gap,(2)}}^{\text{T}}$, respectively. Also, the transport coefficients are specified as $\eta_1 + \eta_2 \approx \frac{4}{3} \alpha \tau$ and $\eta_3 \approx \alpha \tau$ up to linear order in $E$~\cite{Brattan:2024dfv}.\\

To summarize, the dispersion relations derived here unveil rich dynamical electromagnetic behavior in electrically driven nonequilibrium systems. Notably:
\begin{itemize}
\item{Finite electromagnetic coupling modifies both drift and damping of charge fluctuations.}
\item{Distinct electromagnetic modes emerge, converting propagating waves into diffusive or relaxational modes.}
\item{Drift velocities in different channels exhibit measurable splitting.}
\end{itemize}
These findings serve as critical predictions of the extended hydrodynamic model and will be compared with quasi-normal mode spectra from the probe brane analysis in Section~\ref{sec3} and \ref{sec4}.

%
\section{Holographic D3/D7 model with dynamical boundary gauge fields}\label{sec3}
Here, we present the holographic construction as an effective and robust computational framework for analyzing nonequilibrium steady states, providing a novel test of the hydrodynamic theory developed in the previous section. To this end, we employ a probe brane setup~\cite{Karch:2002sh,Karch:2007pd}. 

We begin by reviewing the probe brane system at finite charge density and stationary current, and then introduce the dynamics of linearized fluctuations, focusing on the scenario in which the gauge field is treated dynamically in the boundary field theory through mixed boundary conditions. This setup allows us to capture the essential physics of electrically driven nonequilibrium steady states and study the full dynamical response even beyond the hydrodynamic limit.

%
\subsection{Holographic construction: a quick review} \label{sec:setup}
Probe brane constructions~\cite{Karch:2002sh,Karch:2007pd} offer a concrete realization of electrically driven nonequilibrium steady states, particularly in the regime where the thermal bath is significantly larger than the driven system, enabling strong dissipation.

\paragraph{Probe brane solution.} We consider a D7-brane embedded in a ten-dimensional AdS$_5$-Schwarzschild background with an $S^5$ compactification. The background metric is given by
\begin{equation}
	\dd s^{2} = \frac{\ell^{2}}{u^{2}}\left( -f(u)\dd t^{2} + \dd\vec{x}^{2} +\frac{\dd u^{2}}{f(u)} \right) + \ell^{2}\dd\Omega_{5}^{2}, 
\end{equation}
where $f(u) = 1 - u^4/u_H^4$ is the blackening factor, $\ell$ is the AdS radius, and $u$ denotes the radial coordinate from the AdS boundary ($u = 0$) to the black hole horizon ($u = u_H$). The temperature of the thermal bath is identified with the Hawking temperature, $T = 1/(\pi u_H)$.

In the probe limit, where the D7-brane does not backreact on the background geometry, the dynamics are governed by the Dirac-Born-Infeld (DBI) action
\begin{equation}
	S_{D7} = -T_{D7} \int \dd^{8}\xi \sqrt{-\det \left( g_{ab} + 2\pi\alpha' F_{ab} \right)},
\end{equation}
where $g_{ab}$ is the induced metric on the brane, $\alpha'$ the string coupling, $F_{ab}$ the worldvolume gauge field strength as $\partial_{a}A_{b}-\partial_{b}A_{a}$, and $T_{\text{D7}}$ the brane tension. Without loss of generality, we do not consider the Wess-Zumino term in this paper and focus on trivial embeddings functions, simplifying the internal $S^5$ structure. See \cite{Karch:2002sh,Karch:2007pd} for the details.

\paragraph{Charge and current density.}
To induce a finite charge density and electric field, we adopt the ansatz for our $U(1)$ gauge field
\begin{align}\label{Aansatz}
    A_{t} = A_{t}(u), \qquad A_{x} = -E t + h(u), 
\end{align}
which corresponds to an external electric field $E$ in the $x$-direction. Near the AdS boundary, the gauge field asymptotics yield
\begin{equation}
    A_{t}(u) = \mu -\frac{\rho}{2} u^{2} + \cdots, \qquad
    h(u) = \frac{J}{2}u^{2} + \cdots,
\end{equation}
with $\mu$ as chemical potential, $\rho$ the charge density, and $J$ the induced current. Since the DBI action depends only on $u$-derivatives of the gauge field, the system possesses radially conserved currents~\cite{Karch:2007pd}:
\begin{align}\label{eq:rhoJ}
\rho = \frac{  A_{t}'}{u \sqrt{1 -u^{4} (A_{t}'^{2} +f^{-1}E^{2} -f h'^{2}) }} \,, \quad
  J = \frac{  f h'}{u \sqrt{1 -u^{4} (A_{t}'^{2} +f^{-1}E^{2} -f h'^{2}) }} \,.
\end{align}

These expressions are subject to the condition that the DBI action remains real and positive~\cite{Karch:2007pd}. With the ansatz \eqref{Aansatz}, the DBI action can be further written as
\begin{equation}
	S_{D7} = {\cal{N}} V_{4} \int \dd u g_{xx}\sqrt{-g_{tt}g_{xx}g_{uu} - C^{2}\left(g_{xx}A_{t}'^{2} +g_{uu}E^{2} +g_{tt}h'^{2}\right)}, \label{eq:DBI2}
\end{equation}
where $V_{4} = \int \dd t \dd^{3} x \,,\, {\cal{N}} = T_{D7} (2\pi^{2}) \ell^{8} = {N_{c}}/{\lambda_{\rm t}} \,,\, C = {2\pi \alpha'}/{\ell^{2}}$
with the 't~Hooft coupling $\lambda_{\rm t} = 2\pi g_{\rm s}N_{c}$ and the identification $4\pi g_{\rm s} N_{c} \alpha'^{2} = \ell^{4}$. Hereafter, we set $C=1$ and ${\cal{N}}=1$ for simplicity.
The reality condition on the DBI action leads to the emergence of an effective horizon $u = u_*$ outside the event horizon, $0\leq u_{*} \leq u_{\rm H}$, defined by:
\begin{equation}\label{EFFHO}
 u_{*} = \frac{u_{H}}{\left(1+u_{H}^4E^2\right)^{1/4}} = \frac{1}{\left(E^2 + \pi^4 T^4\right)^{1/4}} \,,
\end{equation}
which sets the causal boundary for dynamics on the D7-brane~\cite{Kim:2011qh,Seiberg:1999vs}, indicating the existence of a nonequilibrium steady state in the dual theory \cite{Kundu:2013eba}.

\paragraph{Nonlinear DC conductivity.}
Given the existence of the effective horizon, we obtain a direct relation between $\rho$ and $J$ as
\begin{equation}
    J^{2} = f(u_{*})\left( \rho^{2} + \frac{1}{u_{*}^{6}}\right) = E^2 u_{*}^4 \left( \rho^{2} + \frac{1}{u_{*}^{6}}\right). \label{eq:J2}
\end{equation}
By applying a constant external electric field to the brane, one induces a stationary current, with dissipation occurring efficiently into the gravitational thermal bath, thus generating a nonequilibrium steady-state configuration.

Consequently, plugging the expression for the effective horizon \eqref{EFFHO} into the definition of the current $J$ \eqref{eq:J2}, a closed-form expression for the nonlinear DC conductivity in the dual field theory is given
\begin{equation}
    \sigma_{\rm DC}\equiv \frac{J}{E} = \pi T \left( \frac{\tilde{\rho}^{2}}{1+ \tilde{E}^{2}} + \sqrt{1+\tilde{E}^{2}} \right)^{1/2},
\end{equation}
with dimensionless variables $\tilde{E} = E / (\pi T)^2$ and $\tilde{\rho} = \rho / (\pi T)^3$. Thus, all physical observables are determined by the two parameters $\tilde{E}$ and $\tilde{\rho}$.

\subsection{Fluctuations and dynamical gauge fields}
To study the linear response, quasi-normal mode spectrum, we introduce fluctuations $A_{\mu}(u) + \delta A_\mu(t,x,y,u)$ around the background solution discussed above. In radial gauge $A_u = 0$ and Fourier space, we define
\begin{equation}
    \delta A_{\mu} = a_{\mu}(u) e^{-i \omega t + i k x} \,,
\end{equation}
and construct two gauge-invariant combinations for computational convenience
\begin{equation}
    Z_{1} \equiv \omega a_{x} + k a_{t}, \quad Z_{2} \equiv a_{y},
\end{equation}
which construct two decoupled fluctuation equations. Here, we assumed that the fluctuations are independent of $z$ owing to the rotational symmetry in the $(y,z)$ plane and the wave-vector is parallel to the $x$-direction. 

\paragraph{Boundary conditions at the (event/effective) horizon.}
To compute the quasi-normal modes, which correspond to the poles of the ``retarded" Green's function, we must impose appropriate boundary conditions both at the black hole horizon and at the AdS boundary.

At the horizon, we impose ingoing wave boundary conditions as prescribed in \cite{Son:2002sd}. In equilibrium (i.e., when the external electric field $E=0$), this condition is implemented at the black hole horizon $u=u_{\rm H}$, and the gauge-invariant field $Z_{\alpha} $ behaves as
\begin{equation}
    Z_{\alpha} = (u-u_{\rm H})^{-i\frac{\omega}{4\pi T}} \tilde{Z}_{\alpha},
\end{equation}
where $\tilde{Z}_{\alpha}$ is a regular function at the horizon. On the other hand, in a non-equilibrium steady state (i.e., when $E\neq 0$), this condition must be modified to reflect the presence of an effective horizon at $u=u_{*}$. The modified form becomes:
\begin{equation}
    Z_{\alpha} = (u-u_{*})^{i \zeta} \bar{Z}_{\alpha},
\end{equation}
where $\bar{Z}_{\alpha}$ is regular at $u=u_{*}$. Setting $\zeta =0$ corresponds to the ingoing wave condition~\cite{Mas:2009wf,Ishigaki:2021vyv}.

\paragraph{Boundary conditions at the AdS boundary and dynamical gauge fields.}
To analyze the quasi-normal modes using the determinant method \cite{Kaminski:2009dh}, we must study the asymptotic behavior of the gauge-invariant fields near the AdS boundary. The expansion of $Z_{\alpha}$ near $u=0$ takes the form as
\begin{equation}
    Z_{\alpha} = Z^{(L)}_{\alpha}\left(1+ \frac{k^{2}-\omega^{2}}{2}u^{2} \log u \right) + Z^{(S)}_{\alpha} u^{2} + \cdots \,, \label{eq:gauge-inv-field-exp}
\end{equation}
where $Z^{(L)}$ and $Z^{(S)}$ are the leading and subleading coefficients, respectively.

In the case of a dynamical gauge field, the correct identification of ``sources" requires a boundary action added to the bulk action~\cite{Ahn:2022azl}. This boundary term is
\begin{equation}\label{BDRYAC}
    S_{\rm bdry} = \int \dd^{4}x \sqrt{-\gamma}\left[-\frac{1}{4\lambda} F_{\mu\nu}F^{\mu\nu} + A_{\mu} J_{\rm ext}^{\mu} \right] + S_{\rm ct},
\end{equation}
where $\gamma$ is the determinant of the induced metric at the boundary cutoff surface $u=\epsilon$, the parameter $\lambda$ characterizes the electromagnetic coupling strength, and $S_{\rm ct}$  is a counterterm added to cancel logarithmic divergences:
\begin{equation}
    S_{\rm ct} = -\frac{\log \epsilon}{4}\int \dd^{4} x \sqrt{-\gamma} F_{\mu\nu}F^{\mu\nu} +{\rm const.}\,
\end{equation}
The constant term includes finite contributions to fix the renormalization scheme. Although this choice affects the Green's function \cite{Mauri:2018pzq}, it does not influence the dispersion of the QNMs. 

Then, varying the total (on-shell) action leads to the boundary Maxwell equation
 \begin{equation}
     \Pi^{\mu} - \frac{1}{\lambda} \partial_{\nu} F^{\mu\nu} +J^{\mu}_{\rm ext} =0 \,, \quad\text{with}\quad \Pi^{\mu} = \frac{\delta S_{\rm on-shell}}{\delta A_{\mu}} \label{condi}
 \end{equation}
indicating that the boundary gauge field $A_{\mu}$ is dynamical and coupled to an external current $J^{\mu}_{\rm ext}$. The quantity $\Pi^{\mu}$ corresponds to the radially conserved bulk current, which, for the $x$-component, coincides with the current $J$ defined in \eqref{eq:rhoJ}.

Finally, taking the first-order variation of the boundary Maxwell equations and using the asymptotic expansion above, we find
\begin{align}
\begin{split}
    \delta \Pi^{x} &= \frac{2\omega}{\omega^{2}-k^{2}}Z^{(S)}_{1}, \qquad\quad\,\,\,
    \delta\Pi^{y} = 2Z_2^{(S)}-\frac{\omega^2-k^2}{2}Z_2^{(L)}\,,\\
    \delta \left(\frac{1}{\lambda} \partial_{\mu} F^{\mu x} \right) &= \frac{\omega}{\lambda} Z^{(L)}_{1} \,, \qquad
    \delta \left(\frac{1}{\lambda} \partial_{\mu} F^{\mu y} \right) = \frac{\omega^2-k^2}{\lambda} Z^{(L)}_{2} \,,
\end{split}
\end{align}
where \eqref{eq:gauge-inv-field-exp} is used. Combining these results, we can express the external current variations as
\begin{equation}
\label{eq:bcs}
\begin{split}
    \delta J_{\rm ext}^{x} = -\frac{\omega}{\lambda} Z_1^{(L)} - \frac{2\omega}{\omega^{2}-k^{2}} Z_1^{(S)}\,,\quad \,\,
    \delta J_{\rm ext}^{y} =\left(\omega^2-k^2\right)\left(\frac{1}{2}-\frac{1}{\lambda}\right)Z_2^{(L)}-2 Z_2^{(S)} \,.
\end{split}
\end{equation}
These expressions represent mixed boundary conditions involving both leading ($Z^{(L)}_{\alpha}$) and subleading ($Z^{(S)}_{\alpha}$) coefficients. 

To identify quasi-normal modes when the gauge fields are dynamical, then we impose that the source terms vanish, i.e.,
\begin{equation}
\begin{split}
\mathcal{S} = \delta J_{\rm ext}^{x, y}=0 \,.
\end{split}
\end{equation}
Since we are working in the probe limit, this condition applies to individual components rather than the full matrix structure. These boundary conditions are essential for correctly computing the spectrum of quasi-normal modes, comparable with the dispersion relations from our hydrodynamic theories.

%
\section{Results at finite electromagnetic coupling}\label{sec4}
Following the method outlined above, we are now prepared to compute the low-energy excitations, quasi-normal modes, within our holographic probe brane model. The phase space of this system is characterized by three scale-invariant parameters:
\begin{equation}
\{\rho/\left(\pi T\right)^3,\, E/\left(\pi T\right)^2,\, \lambda\} \,.
\end{equation}
In this work, we primarily focus on the regime of large charge density, specifically
\begin{equation}
\rho/(\pi T)^{3}=10^{5} \gg 1 \,,
\end{equation}
where the framework of relaxed hydrodynamics is valid, as demonstrated in Refs.~\cite{Chen:2017dsy, Brattan:2024dfv}.
Note that this regime is discussed in terms of the so-called holographic zero sound mode \cite{Davison:2011ek} (see also \cite{Karch:2009zz}).

In this large density limit, several key parameters can be obtained analytically~\cite{Chen:2017dsy}. They are given by:
\begin{equation}
\label{eq:vchitau}
v^{2} = \frac{1}{3},  \qquad \frac{1}{\tau} =\frac{(\pi T)^{2}}{c \rho^{1/3}}, \qquad \chi =\frac{\rho^{2/3}}{c},
\end{equation}
with $c= {\Gamma(1/3)\Gamma(1/6)}/{(6\sqrt{\pi})}$ and other transport coefficients ($\alpha$, $\eta_{i}$) are numerically determined as~\cite{Brattan:2024dfv}
\begin{equation}
\alpha \approx \frac{1}{\tau}\,,\qquad \eta_1+\eta_2 \approx \frac{4}{3}\,, \qquad \eta_3 \approx 1 \,,
\end{equation}
valid to linear order in $E$. These coefficients are subject to corrections of order $O(E^{2})$; nevertheless, in the remainder of this work, we mainly focus on the regime of small $E$.

In what follows, we analyze the role of the electromagnetic coupling parameter $\lambda$, and examine how its variation influences the quasi-normal modes. To this end, we consider two distinct cases: zero electric field $(E/\left(\pi T\right)^2=0)$ and finite electric field $(E/\left(\pi T\right)^2\neq0)$, which are discussed separately.

\subsection{Warm-up: equilibrium states ($E=0$)}
\begin{figure}[]
    \centering
    \includegraphics[width=0.45\textwidth]{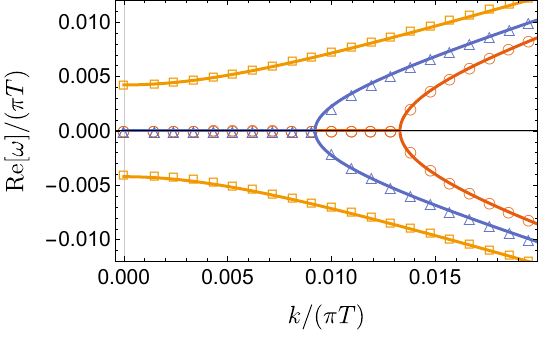}
    \includegraphics[width=0.45\textwidth]{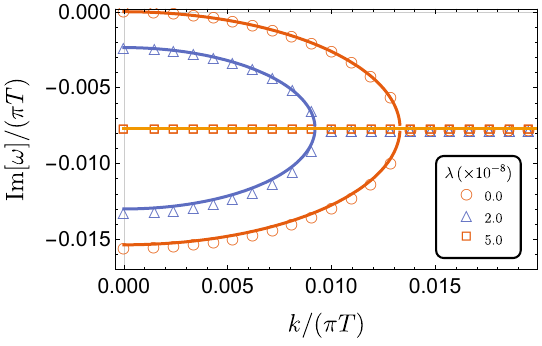}
	\caption{Dispersion relations of the lowest quasi-normal modes of longitudinal sector at zero electric field $(E/\left(\pi T\right)^2=0$) with $\lambda = (0, 2, 5) \times 10^{-8}$. Symbols are the numerical data, while solid lines the theoretical predictions from Eq.~\eqref{eq:gapE0eq}.}
    \label{fig:dispersionE0}
\end{figure}

We begin our analysis by considering the equilibrium state characterized by a zero electric field, $E=0$. Specifically, our focus is on the longitudinal sector described by equation \eqref{eq:hydrolongi}, which simplifies under this condition to
\begin{equation}
    \omega \left( \omega + \frac{i}{\tau}\right) = v^{2}k^{2} + \lambda \chi \,,
\label{eq:gapE0eq}
\end{equation}
where we used \eqref{eq:c2}. Numerical results for quasi-normal modes match well with the dispersion relations derived from hydrodynamics as expressed by equations \eqref{eq:gapE0eq}. This agreement is clearly illustrated in Figure \ref{fig:dispersionE0}.

Figure \ref{fig:dispersionE0} demonstrates the behavior of the lowest quasi-normal modes for various electromagnetic couplings ($\lambda$). Notably, in the limit of very small $\lambda$, two distinct non-hydrodynamic poles emerge, one corresponding explicitly to the damped charge diffusion mode. As the electromagnetic coupling $\lambda$ increases, these poles move closer together along the negative imaginary frequency axis, ultimately approaching each other.
\begin{figure}[]
    \centering
    \includegraphics[width=0.45\textwidth]{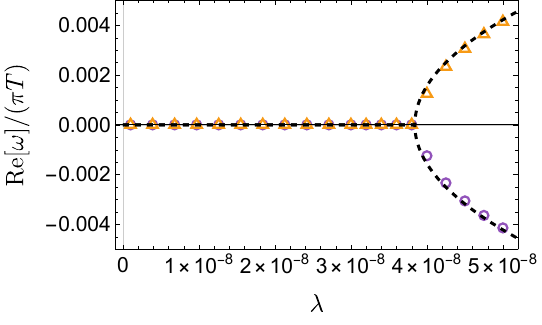}
    \includegraphics[width=0.45\textwidth]{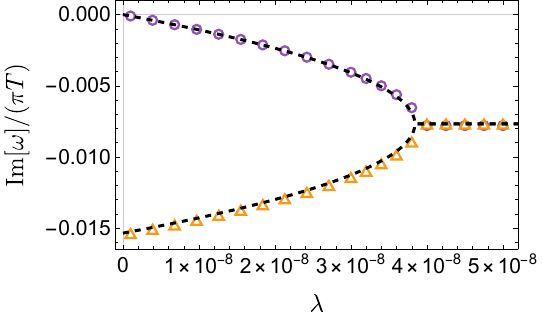}
    \caption{Non-hydrodynamic modes of longitudinal sector at zero electric field $(E/\left(\pi T\right)^2=0$). Symbols are the numerical data, while solid lines the theoretical predictions from Eq.~\eqref{eq:gapE0}. The wave-vector is set to zero here.}
    \label{fig:lambdadep}
\end{figure}

As such, by the effect of $\lambda$, the two lowest non-hydrodynamic modes acquire a finite real gap and identical imaginary dispersion, given explicitly by
\begin{equation}
    \omega(k=0) = \pm \sqrt{\lambda \chi - \frac{1}{4\tau^{2}}}  -\frac{i}{2\tau} \,.
    \label{eq:gapE0}
\end{equation}
This behavior, characterized by the presence of a finite real gap, resembles the emergence of plasma modes previously observed at large density regimes under finite coupling $\lambda$~\cite{Ahn:2022azl}. The numerical validation of this non-hydrodynamic mode behavior is shown explicitly in Figure \ref{fig:lambdadep} (see also Appendix.~\ref{sec:cccf}).

\subsection{Non-equilibrium steady states with electromagnetic couplings}
Next, we explore the dispersion relations at a finite electric field, characterized by the condition $E/(\pi T)^2 \neq 0$. From the hydrodynamic analysis presented in Section \ref{sec2}, we expect two longitudinal modes, as described in Eq.~\eqref{LONGIDIS}, and three transverse modes, outlined in \eqref{TRANIDIS}, emerging in the limit of small wave-vectors. Here, we confirm that the computed quasi-normal modes align precisely with these predictions, especially matching the complete dispersion relations provided by hydrodynamics for both longitudinal \eqref{eq:hydrolongi} and transverse modes \eqref{eq:hydrotrans}.

\begin{figure}[]
    \centering
    \includegraphics[width=0.45\textwidth]{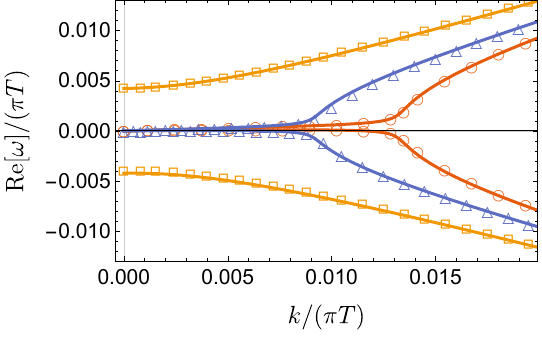}
    \includegraphics[width=0.45\textwidth]{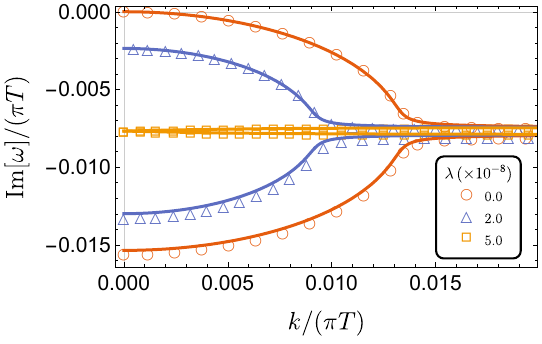}
	\caption{Dispersion relations of the lowest quasi-normal modes of longitudinal sector at finite electric field $(E/\left(\pi T\right)^2=0.05$) for different values of $\lambda = (0, 2, 5) \times 10^{-8}$. Symbols are the numerical data while solid lines the theoretical predictions from Eq. \eqref{eq:hydrolongi}.}
    \label{fig:dispersion}
\end{figure}

Figure \ref{fig:dispersion} illustrates the lowest quasi-normal modes within the longitudinal sector at finite electric field. The results exhibit excellent agreement with the hydrodynamic dispersion relation in Eq. \eqref{eq:hydrolongi}.
By comparing Figure \ref{fig:dispersion} with the corresponding scenario at zero electric field (Figure \ref{fig:dispersionE0}), we identify two clear effects induced by the electric field:
\begin{itemize}
\item{The real poles exhibit a finite drift velocity, corresponding to a linear-in-$k$ term in the dispersion relation, Eq. \eqref{LONGIDIS}.}
\item{In the large wave-vector regime, the imaginary poles split into two distinct branches.}
\end{itemize}

\begin{figure}[tbp]
    \centering
    \includegraphics[width=0.45\textwidth]{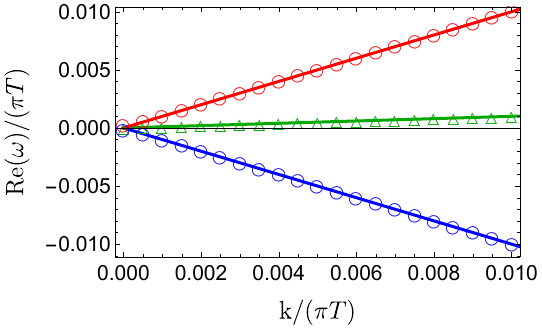}
    \includegraphics[width=0.46\textwidth]{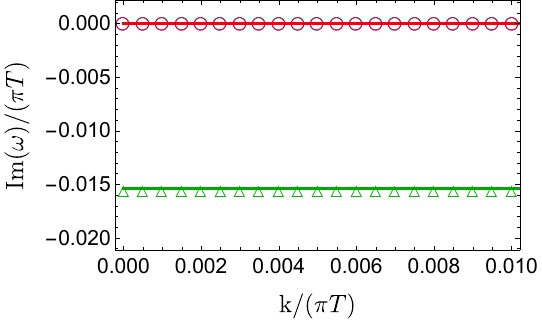}

    \includegraphics[width=0.45\textwidth]{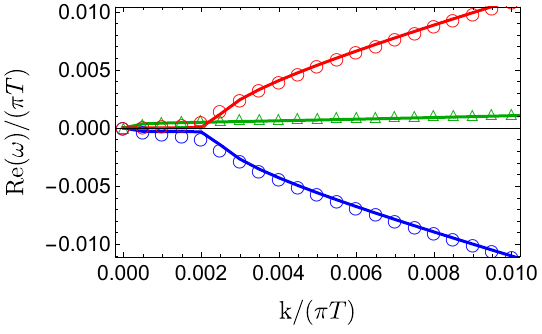}
    \includegraphics[width=0.45\textwidth]{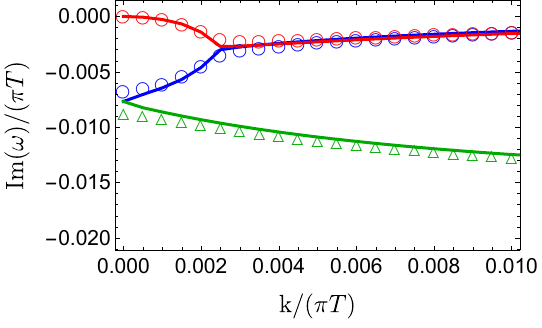}
    \caption{Dispersion relations of the lowest quasi-normal modes of transverse sector at finite electric field $(E/\left(\pi T\right)^2=0.1$). The upper and lower panels represent the scenarios with $\lambda = 0$ and $\lambda = 3.8 \times 10^{-8}$, respectively. Symbols are the numerical data while solid lines the theoretical predictions from Eq. \eqref{eq:hydrotrans}.}
    \label{fig:Tlam0}
\end{figure}

In Figure \ref{fig:Tlam0}, we also present the quasi-normal modes of the transverse sector at finite electric field, demonstrating precise consistency with the hydrodynamic dispersion relations given by \eqref{eq:hydrotrans}. We notice that the zero-electric-field limit ($E=0$) within the transverse sector can be readily understood from \eqref{TRANIDIS}.

For the case of $\lambda=0$ (upper panel of Figure \ref{fig:Tlam0}), we observe propagating photon modes (indicated by red and blue lines) with dispersion relations given by $\omega = \pm k$. Additionally, we identify the mode at finite drift velocity, labeled $\omega_{\text{gap, (2)}}^{\text{T}}$ in Eq.\eqref{TRANIDIS}, indicated in green.

When introducing finite $\lambda$ (lower panel of Figure \ref{fig:Tlam0}), the propagating photons are modified into two distinct dispersion relations: the gapless transverse mode $\omega_{\text{gapless}}^{\text{T}}$ (red line), and the gapped mode $\omega_{\text{gap, (1)}}^{\text{T}}$ (blue line) in \eqref{TRANIDIS}. The gapped mode features a negative linear-in-$k$ term in the real part of the dispersion, along with a quadratic dependence in its imaginary part. In this specific scenario, two imaginary gap modes (blue and green lines) overlap precisely due to the chosen value of $\lambda=3.8 \times 10^{-8}$.

\subsection{On the electric field dependence: screening length and relaxation time}
Lastly, we conclude this section by discussing the effect of electric fields on the dispersion relations, especially with two special cases: (I) the static case $(\omega = 0)$ and (II) the homogeneous case $(k = 0)$.

\paragraph{Static case: screening lengths.}
We first analyze the static scenario $(\omega = 0)$ within the longitudinal mode framework. Considering a perturbation solely along $\delta J_x$ with the electric field aligned as $\vec{E}=(E,0,0)$, the linearized charge fluctuation equation (\ref{eq:chargefluc}) simplifies to:
\begin{equation}
	\left(   D \partial_x^2 -E  \partial_x   - K  \right) \delta \rho(x) =0, \label{eq:drift-diffusion-decay}
\end{equation}
where we used $\alpha\tau \simeq 1$, valid to first order in $E$ \cite{Brattan:2024dfv}. Here, each term represents a distinct physical phenomenon: diffusion, advection, and decay of the charge fluctuations, respectively.

Solving this equation in Fourier space, we obtain two characteristic solutions
\begin{equation}
	k_\pm = i \frac{E \pm \sqrt{E^{2}+4DK}}{2D}, \label{eq:kpm}
\end{equation}
which define two distinct screening lengths, given by $\lambda_{\rm s}^{\pm} = |k_{\pm}|^{-1}$.

These two screening lengths differ in the steady-state scenario due to the presence of electromagnetic coupling $(\lambda \neq 0, K \neq 0)$. Notably, when the coupling $\lambda$ vanishes, we recover a single screening length, consistent with previous analyses \cite{Brattan:2024dfv}. Additionally, the limit $E\rightarrow 0$ merges these two scales into the familiar Debye screening length, $\lambda_{\rm D}=\sqrt{K/D}$.

\begin{figure}[tbp]
    \centering
    \includegraphics[width=0.7\textwidth]{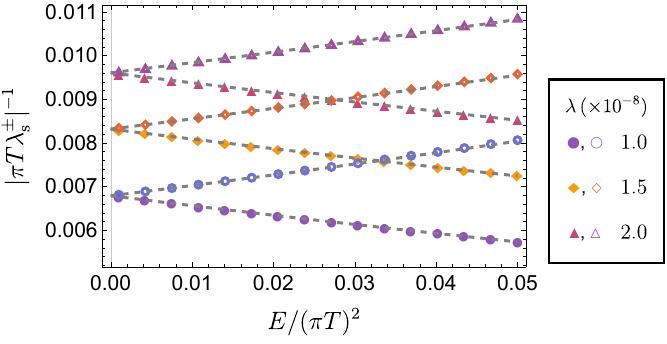}
	\caption{Inverse screening lengths $\lambda_{\rm s}^{\pm} = |k_{\pm}|^{-1}$ as functions of the electric field $E$. The analytical prediction from equation (\ref{eq:kpm}) is represented by gray dashed lines.}
    \label{fig:omegazero}
\end{figure}

Our hydrodynamic predictions from Eq. (\ref{eq:kpm}) show excellent agreement with numerical data derived from quasi-normal modes. Both screening lengths exhibit monotonic behavior (either increasing or decreasing) as the electric field magnitude changes: Figure \ref{fig:omegazero} displays the inverse screening lengths $\lambda_{\rm s}^{\pm} = |k_{\pm}|^{-1}$ as a function of the electric field $E$. 

Specifically, within the D3/D7 model, we determine these screening lengths by locating poles in the Green’s function $G_{\rho\rho}(\omega=0,k)$, characterized by zeros of the mixed boundary condition for the static limit
\begin{equation}
	\delta J_\mathrm{ext}^{t} = -\frac{k^2}{\lambda} a_t^{(0)} + 2 a_t^{(2)}.
\end{equation}
As evident from figure \ref{fig:omegazero}, the numerical values for the inverse screening lengths, obtained from the Green's function poles, closely follow the analytical predictions, confirming the validity of our hydrodynamic analysis.

\paragraph{Homogeneous case: relaxation times.}
We now examine the depencence of the homogeneous case ($k=0$), gapped modes, on the electric field $E$. A comparision between Figure \ref{fig:dispersionE0} and Figure \ref{fig:dispersion} reveals the following behavior. In the case of $\lambda=0$, the hydrodynamic mode remains at $\omega=0$, while the non-hydrodynamic mode shifts deeper into the complex plane, with its damping rate characterzied by $\tau^{-1}(E)$. In contrast, when $\lambda \neq 0$, both modes acquire a finite imaginary gap and progressively approach each other as $\lambda$ increases. Notably, the electric field $E$ affacts the two modes in opposing ways, effectively counteracting the behavior induced by electrodynamic coupling $\lambda$.

\begin{figure}[tbp]
    \centering
    \includegraphics[width=0.45\textwidth]{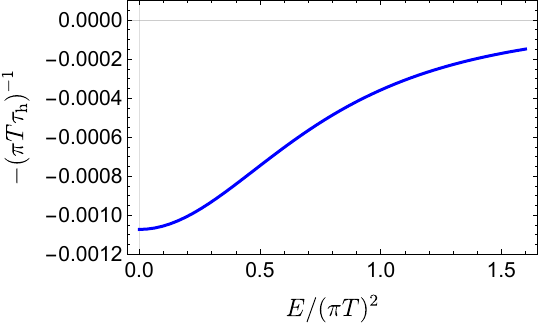}
    \includegraphics[width=0.44\textwidth]{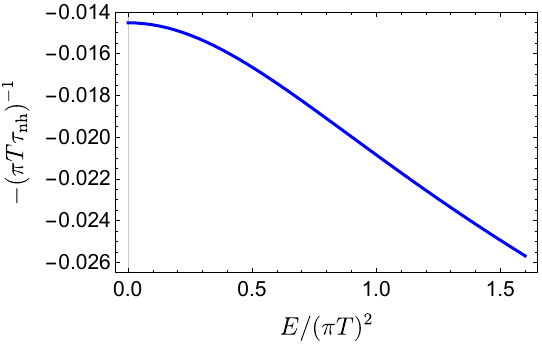}
	\caption{The $E$ dependence of the upper imaginary gap (left) and lower imaginary gap (right) when $\lambda=10^{-8}$.}
    \label{fig:imgap}
\end{figure}

Figure \ref{fig:imgap} shows such a $E$-dependence of the relaxation times associated with the two modes, denoted by $\tau_{\rm h}$ and $\tau_{\rm nh}$, at finite $\lambda$.
As shown, the upper imaginary gap ($\sim \tau_{\text{h}}^{-1}$) decreases with increasing $E$, moving closer to the origin, whereas the lower imaginary gap ($\sim \tau_{\text{nh}}^{-1}$) increases with $E$, moving deeper into the complex plane.

%
\section{Conclusion}
Nonequilibrium steady states (NESS) represent systems that, while not in equilibrium, exhibit steady, time-independent macroscopic behavior. Such states arise naturally when external driving forces balance internal dissipation, maintaining continuous energy or particle currents. Examples include electrical conductivity described by the Drude model, and thermal conduction along a metal rod. Nevertheless, understanding these steady yet nonequilibrium conditions poses a substantial theoretical challenge: standard equilibrium thermodynamics and hydrodynamics need extensions to encompass NESS accurately.

In this work, we have developed an extended hydrodynamic framework for describing NESS in charged systems, explicitly incorporating dynamical electromagnetic fields. This extension addresses a crucial gap in conventional hydrodynamics by enabling the consistent treatment of electromagnetic interactions within systems driven by external electric fields.

To validate our theoretical formulation, we employed holographic probe brane constructions based on the D3/D7 model. In addition, by implementing mixed boundary conditions on the bulk gauge fields, we effectively promoted the global boundary $U(1)$ symmetry to a dynamical gauge symmetry, thereby introducing electromagnetic fields as active components in the dual field theory. This construction allowed us to compute the spectrum of quasinormal modes (QNMs) and to analyze their behavior in both equilibrium and nonequilibrium regimes.

Our results reveal several rich dynamical behaviors. First, the QNM spectra obtained from the holographic model closely match the dispersion relations predicted by the extended hydrodynamic theory. In particular, the electromagnetic coupling introduces significant modifications to both the damping rates and drift velocities of charge fluctuations. Second, we identified novel collective excitations --electromagnetic modes that interpolate between propagating, diffusive, and relaxational behaviors-- whose presence is a direct consequence of the dynamical gauge fields. Third, our analysis of static and homogeneous limits provided further evidence for electromagnetic screening effects and revealed distinct relaxation timescales associated with different hydrodynamic modes.

Looking ahead, our approach opens several directions for further research. These include studying the interplay between dynamical gauge fields and additional symmetry-breaking effects, exploring non-linear transport phenomena, and extending the analysis to other holographic setups and higher-derivative corrections. Ultimately, we expect this line of inquiry to deepen our understanding of nonequilibrium dynamics in systems relevant to quantum chromodynamics and condensed matter physics.

\acknowledgments
YA and MB acknowledge the support of the Shanghai Municipal Science and Technology Major Project (Grant No.2019SHZDZX01). MB acknowledges the sponsorship from the Yangyang Development Fund.
HSJ is supported by the Spanish MINECO ‘Centro de Excelencia Severo Ochoa' program under grant SEV-2012-0249, the Comunidad de Madrid ‘Atracci\'on de Talento’ program (ATCAM) grant 2020-T1/TIC-20495, the Spanish Research Agency via grants CEX2020-001007-S and PID2021-123017NB-I00, funded by MCIN/AEI/10.13039/501100011033, and ERDF `A way of making Europe.'
%

\appendix
\section{Current-current correlation function}
\label{sec:cccf}
In this appendix, we discuss the current-current correlation function for equilibrium states in the presence of the electromagnetic coupling. In electromagnetism, the electric displacement field is given by\footnote{In the main text, we denote the fluctuations of the fields with $\delta$, but we omit writing here for simplicity.}
\begin{equation}
    \vec{D} = \epsilon \vec{E} = \vec{E}+\vec{P},
\end{equation}
where the polarization field $\vec{P}$ satisfies $\vec{\nabla} \cdot \vec{P} = - \rho/\epsilon_{e} = -\lambda \rho$ and $\epsilon$ is the dielectric function. Using the Maxwell's equation and the conservation equation, the dielectric function is written with the conductivity in momentum space as
\begin{equation}
    \epsilon(\omega,\vec{k}) = 1- \frac{\lambda}{i\omega}\sigma(\omega,\vec{k}).
\end{equation}
If we define the conductivity including the effect of the electromagnetic coupling as $\vec{J} = \sigma_{\rm em} \vec{D}$, it is written as
\begin{equation}
    \sigma_{\rm em}(\omega, \vec{k}) = \frac{\sigma(\omega, \vec{k})}{1- \frac{\lambda}{i \omega} \sigma(\omega, \vec{k})}.
\end{equation}
The optical conductivity can be computed by the retarded current-current correlation function $\chi_{JJ}(\omega, \vec{k}=0) = i \omega \sigma_{\rm em}(\omega)$, that is
\begin{equation}
    \chi_{JJ}(\omega) =  \frac{G_{JJ}^{R}(\omega)} {1+\frac{\lambda}{\omega^{2}}G_{JJ}^{R}(\omega)},
\end{equation}
where $G_{JJ}^{R}(\omega,\vec{k}=0)=i \omega \sigma(\omega,\vec{k}=0) $ is the retarded Green's function for the longitudinal current that is derived by the usual holographic prescription \cite{Son:2002sd}. 

\begin{figure}[tbp]
    \centering
    \includegraphics[width=0.45\textwidth]{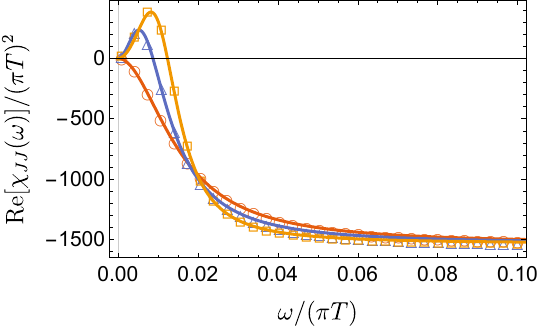}
    \includegraphics[width=0.45\textwidth]{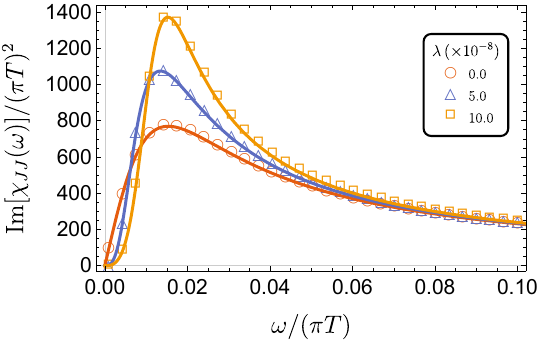}
	\caption{The real (left) and imaginary (right) part of the current-current correlation function for several values of $\lambda$. The analytical formula \eqref{eq:current-current-correlation} is denoted by the solid lines.}
    \label{fig:correlation}
\end{figure}

From the constitutive relation \eqref{eq:currentflucdyn} with $E^{i}=0$ and the charge conservation equation, the optical conductivity is given by the Drude form:
\begin{equation}
    \sigma(\omega) = \frac{\sigma_{\rm DC}}{1- i \tau\omega},
\end{equation}
with $\sigma_{\rm DC} = \chi \tau$, and we obtain
\begin{equation}
    \chi_{JJ}(\omega) =  \frac{i \sigma_{\rm DC} \omega^{2}}{\omega(1-i\tau \omega) + i \lambda \sigma_{\rm DC}}.
    \label{eq:current-current-correlation}
\end{equation}
In figure \ref{fig:correlation}, we show the comparison between this analytical form and the numerical results by the holographic D3/D7 model, and confirm their good agreement.

\begin{figure}[tbp]
    \centering
    \includegraphics[width=0.7\textwidth]{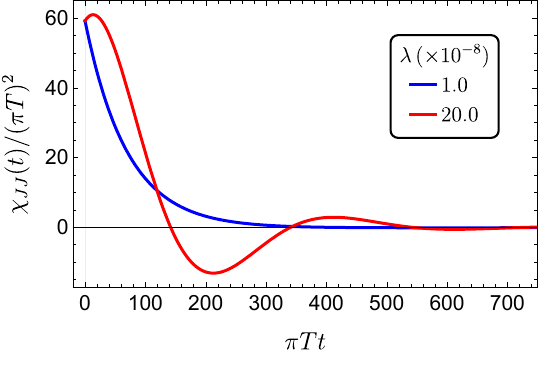}
	\caption{The current-current correlation function as a function of time for two different values of $\lambda$.}
    \label{fig:responseT}
\end{figure}

As a further check on the effect of the electromagnetic coupling, we show the retarded current-current correlation function as a function of time in Figure \ref{fig:responseT} obtained by performing the inverse Fourier transformation of \eqref{eq:current-current-correlation}. It exhibits an overdamped behavior for smaller $\lambda$, while an underdamped oscillating behavior for larger $\lambda$. This is consistent with the emergence of a finite real gap in dispersion relation as shown in Fig. \ref{fig:lambdadep}.

\bibliographystyle{JHEP}
\providecommand{\href}[2]{#2}\begingroup\raggedright\endgroup

\end{document}